\documentclass[conference]{IEEEtran}
\IEEEoverridecommandlockouts
\usepackage{cite}
\usepackage{amsmath,amssymb,amsfonts}
\usepackage{algorithmic}
\usepackage{graphicx}
\usepackage{textcomp}
\usepackage{xcolor}
\def\BibTeX{{\rm B\kern-.05em{\sc i\kern-.025em b}\kern-.08em
    T\kern-.1667em\lower.7ex\hbox{E}\kern-.125emX}}
\begin{document}

\title{Cyber-Resilient Fault Diagnosis Methodology in Inverter-Based Resource-Dominated Microgrids with Single-Point Measurement}

\author{\IEEEauthorblockN{Yifan Wang}
\IEEEauthorblockA{\textit{School of Electrical and Electronic Engineering} \\
\textit{Nanyang Technological University}\\
Singapore, Singapore \\
wang2115@e.ntu.edu.sg}
\and
\IEEEauthorblockN{Yiyao Yu}
\IEEEauthorblockA{\textit{College of Design and Engineering} \\
\textit{National University of Singapore}\\
Singapore, Singapore \\
e1350878@u.nus.edu}
\and
\IEEEauthorblockN{Yang Xia}
\IEEEauthorblockA{\textit{School of Electrical and Electronic Engineering} \\
\textit{Nanyang Technological University}\\
Singapore, Singapore \\
yang\_xia@ntu.edu.sg}
\and
\IEEEauthorblockN{Yan Xu}
\IEEEauthorblockA{\textit{School of Electrical and Electronic Engineering} \\
\textit{Nanyang Technological University}\\
Singapore, Singapore \\
xuyan@ntu.edu.sg}
}

\maketitle

\begin{abstract}
Cyber-attacks jeopardize the safe operation of inverter-based resource-dominated microgrids (IBR-dominated microgrids). At the same time, existing diagnostic methods either depend on expensive multi-point instrumentation or stringent modeling assumptions that are untenable under single-point measurement constraints. This paper proposes a Fractional-Order Memory-Enhanced Attack-Diagnosis Scheme (FO-MADS) that achieves timely fault localization and cyber-resilient fault diagnosis using only one VPQ (voltage, active power, reactive power) measurement point. FO-MADS first constructs a dual fractional-order feature library by jointly applying Caputo and Grünwald-Letnikov derivatives, thereby amplifying micro-perturbations and slow drifts in the VPQ signal. A two-stage hierarchical classifier then pinpoints the affected inverter and isolates the faulty IGBT switch, effectively alleviating class imbalance. Robustness is further strengthened through Progressive Memory-Replay Adversarial Training (PMR-AT), whose attack-aware loss is dynamically re-weighted via Online Hard Example Mining (OHEM) to prioritize the most challenging samples.

Experiments on a four-inverter IBR-dominated microgrid testbed comprising 1 normal and 24 fault classes under four attack scenarios demonstrate diagnostic accuracies of 96.6\% (bias), 94.0\% (noise), 92.8\% (data replacement), and 95.7\% (replay), while sustaining 96.7\% under attack-free conditions. These results establish FO-MADS as a cost-effective and readily deployable solution that markedly enhances the cyber-physical resilience of IBR-dominated microgrids.
\end{abstract}


\section{Introduction}
Modern power systems are experiencing rapid growth in distributed energy resources (DERs) such as solar photovoltaics, wind turbines, and battery storage \cite{ref1}. Inverter-based resource-dominated microgrids (IBR-dominated microgrids) integrate these DERs and, through power electronic converters (inverters), interface renewable generation and storage with the grid \cite{ref1,ref2}. Converter failures can jeopardize IBR-dominated microgrid stability, so continuous monitoring and robust diagnostics are required \cite{ref1,ref2}. Recent studies have proposed data-driven monitoring techniques \cite{ref3,ref4}, yet open-circuit faults in insulated-gate bipolar transistor (IGBT) switches can still evade detection and cause phase current imbalance, torque ripple, and power quality degradation \cite{ref5}.

Conventional converter fault diagnosis is typically model-based or signal-based \cite{ref6}. Model-based schemes rely on accurate system parameters and are sensitive to drift, whereas signal-based methods can be fragile to noise and changes in operating conditions \cite{ref6,ref7}. At the same time, integrating information and communication technology (ICT) exposes IBR-dominated microgrids to cyber vulnerabilities such as false data injection (FDI), denial-of-service (DoS), and deception attacks \cite{ref8,ref9}. These cyber threats complicate diagnosis and have motivated attack-resilient control strategies \cite{ref6,ref9,ref10}.

With the rapid development of artificial intelligence, data-driven fault diagnosis for power converters has become a major trend \cite{ref11,ref12,ref13,ref14}. Machine learning models map sensor data directly to fault categories, enabling rapid detection without explicit physical models \cite{ref11,ref12}. However, their performance may degrade with low-quality data or unseen operating scenarios, especially when they depend on \textit{multi-point} measurements that increase cost and complexity \cite{ref4,ref12}. Moreover, most existing studies still assume attack-free conditions and do not explicitly address concurrent cyber-attacks during fault diagnosis \cite{ref6,ref8,ref15,ref16,ref17,ref18,ref19}.

To fill these research gaps, this paper proposes FO-MADS, a unified framework that is both measurement-efficient and cyber-resilient. FO-MADS detects and localizes inverter open-circuit faults using only a single measurement point (voltage and power at the point of common coupling (PCC)) while simultaneously monitoring for cyber-attacks. Fractional-order signal processing constructs a dual-feature library: the Caputo derivative accentuates high-frequency perturbations for fault detection, whereas the Grünwald-Letnikov derivative emphasizes slow drifts for stealthy cyber-attacks. A two-stage hierarchical classifier localizes the faulty inverter (Stage~1) and pinpoints the specific faulty IGBT switch (Stage~2) \cite{ref12,ref13}, while adversarial training is integrated via PMR-AT to improve robustness under diverse attack scenarios \cite{ref20,ref21,ref22}.

The main contributions are:
(i) \textit{Single-point, dual-fractional feature fault diagnosis}: A framework requiring only one PCC measurement with complementary fractional-order derivatives.
(ii) \textit{Hierarchical fault localization}: Two-stage strategy isolating faulty inverter and pinpointing faulty IGBT switch.
(iii) \textit{Cyber-resilient fault diagnosis}: PMR-AT ensures robust performance under cyber-physical disturbances.

\begin{figure*}[!t]
\centering
\includegraphics[width=\textwidth]{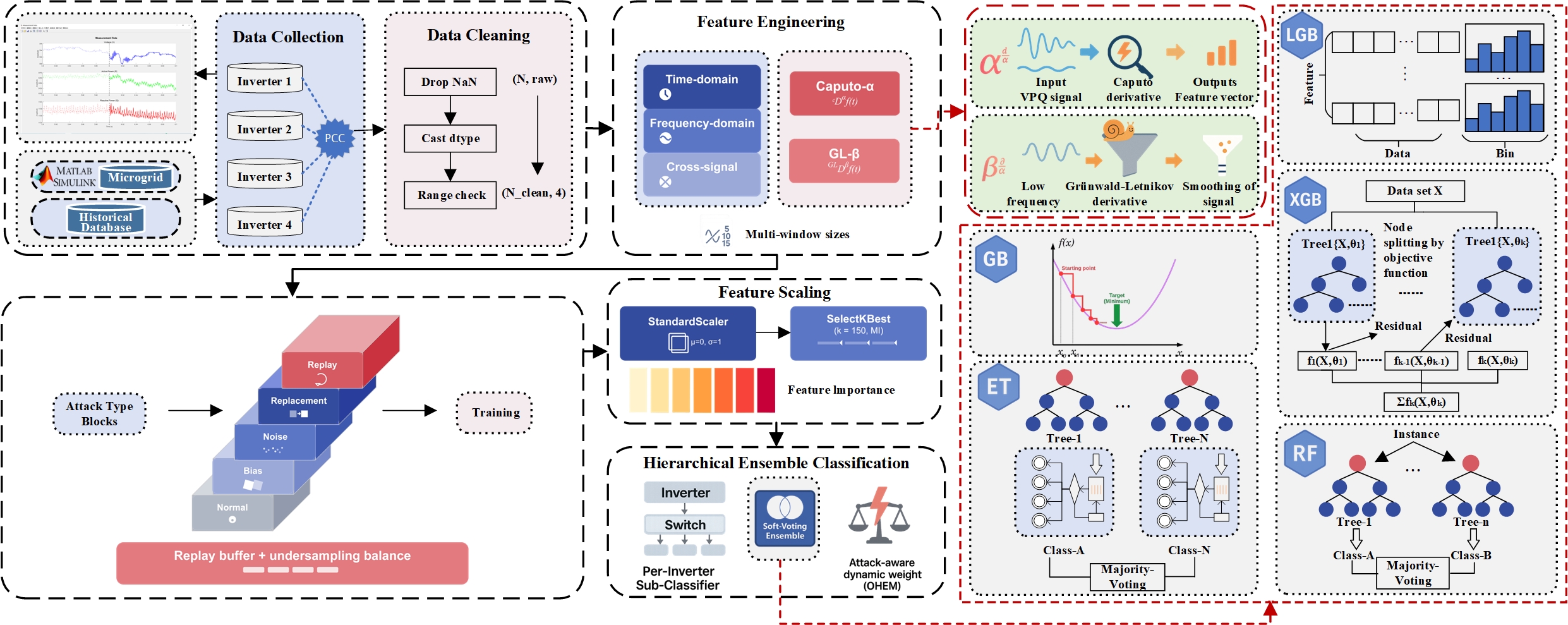}
\caption{Main framework of the FO-MADS}
\label{fig:framework}
\end{figure*}

The proposed FO-MADS bridges signal processing, machine learning, and cyber-security, providing a cost-effective route to improve IBR-dominated microgrid reliability. Effectiveness is validated on a multi-inverter testbed under various faults and cyber-attacks. The overall architecture of the proposed FO-MADS framework is depicted in Fig.~\ref{fig:framework}.

\section{Proposed FO-MADS Framework}
FO-MADS utilizes single-point PCC data ($V$, $P$, $Q$) to classify 25 operational states across four inverters \cite{ref20,ref21,ref22}. As illustrated in Fig.~\ref{fig:framework}, the framework consists of dual-definition fractional-order feature extraction, a hierarchical diagnostic architecture, and adversarially robust training.

\subsection{Dual-Definition Fractional-Order Feature Engineering}
\begin{figure}[!t]
\centering
\includegraphics[width=\linewidth]{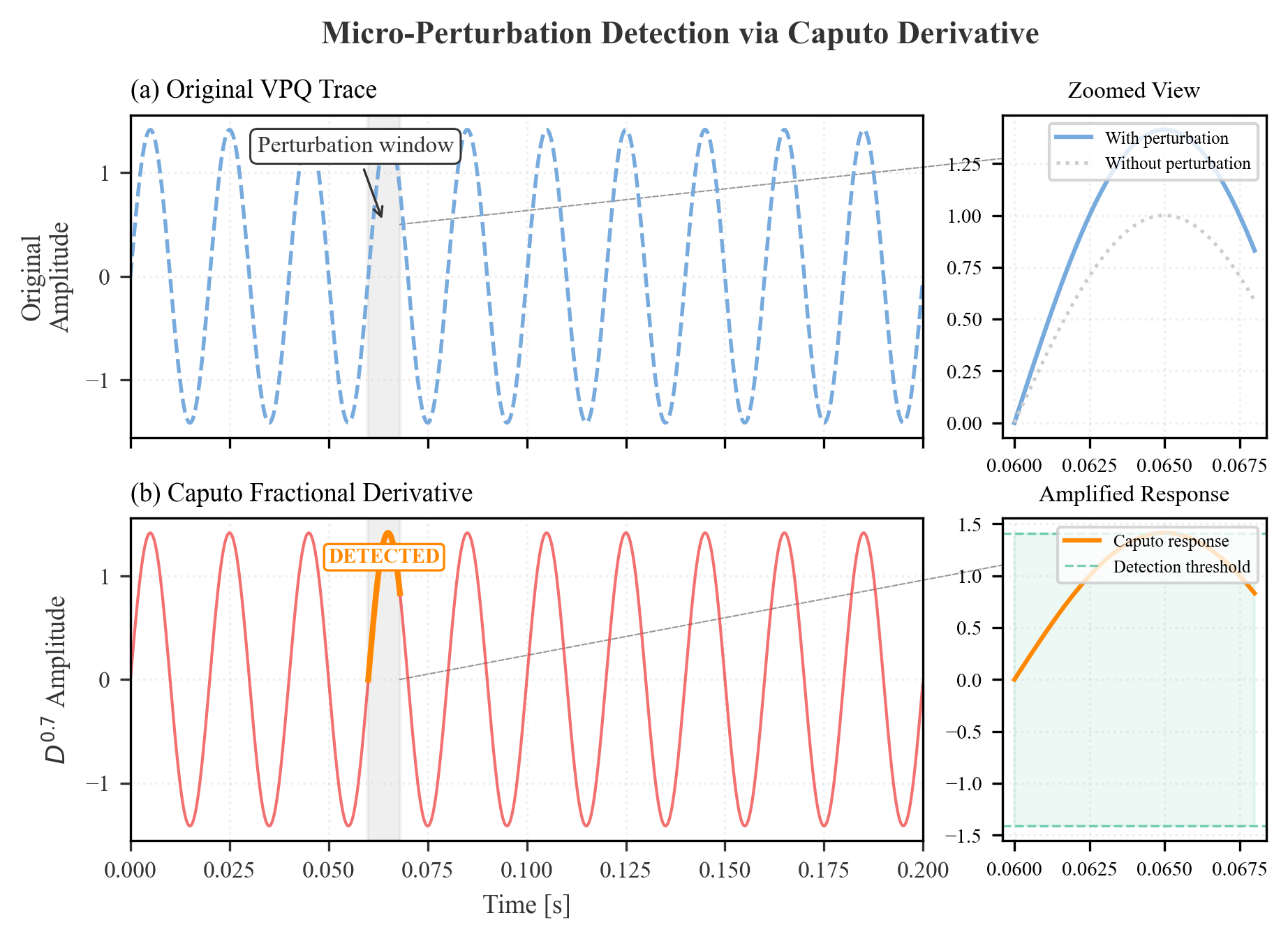}
\caption{Process of Caputo derivative on VPQ signals showing enhanced detection of high-frequency transients}
\label{fig:caputo_effect}
\end{figure}

Two fractional operators capture complementary dynamics \cite{ref23}: 1) \textbf{Caputo Derivative} (micro-perturbation detection):
\begin{equation}
{}^{C}D_{t}^{\alpha} f(t) = \frac{1}{\Gamma(1-\alpha)} \int_0^t (t - \tau)^{-\alpha} f'(\tau) \, d\tau.
\label{eq:caputo}
\end{equation}
As shown in Fig.~\ref{fig:caputo_effect}, the Caputo derivative enhances high-frequency transients in the VPQ signals, making subtle switching faults more separable in the feature space.

\begin{figure}[!t]
\centering
\includegraphics[width=\linewidth]{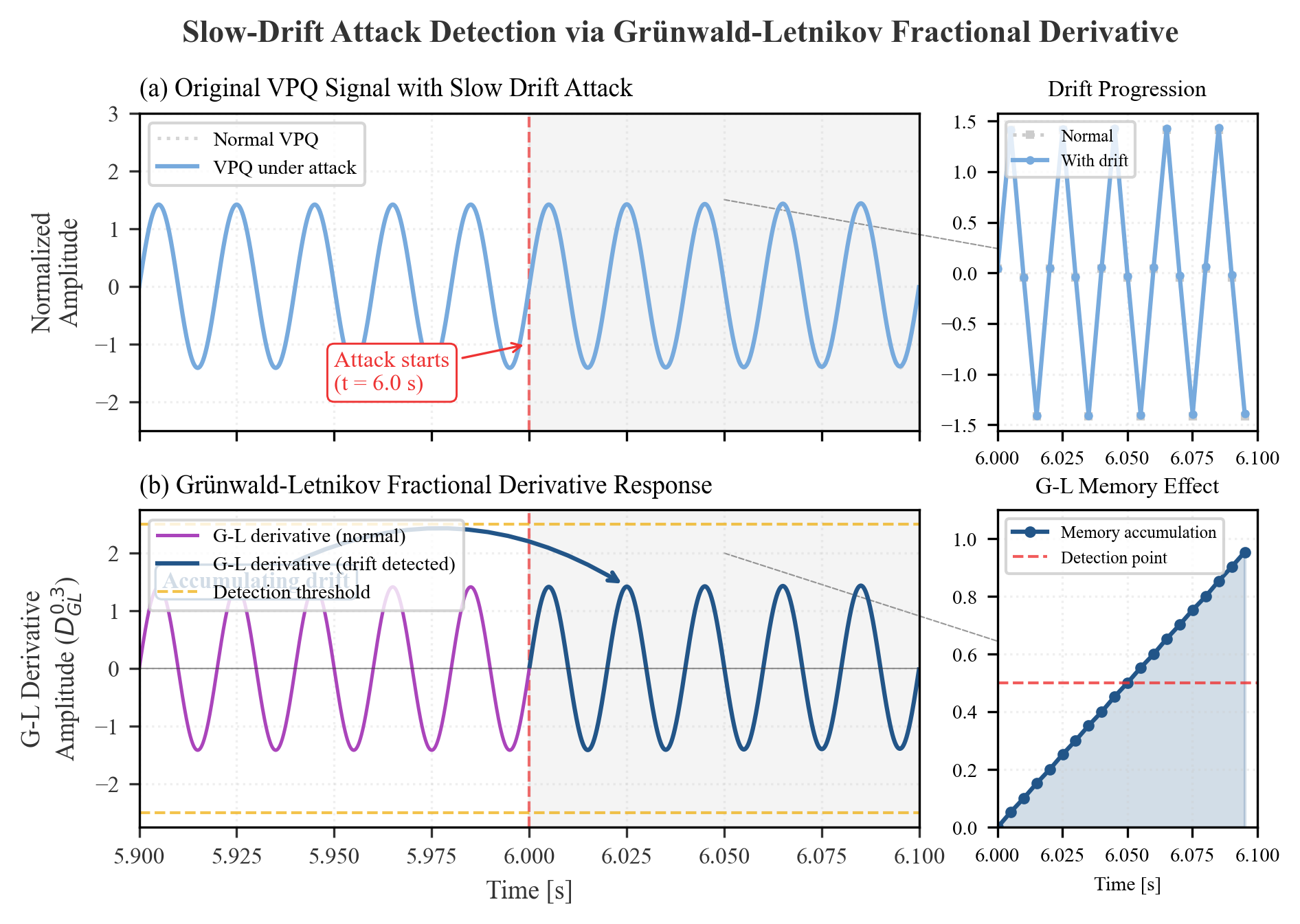}
\caption{Process of Grünwald-Letnikov derivative on VPQ signals showing enhanced detection of slow-drift anomalies}
\label{fig:gl_effect}
\end{figure}

2) \textbf{Grünwald-Letnikov Derivative} (slow-drift detection):
\begin{equation}
{}^{GL}D_{t}^{\beta} f(t) = \lim_{h \to 0} h^{-\beta} \sum_{k=0}^{\lfloor t/h \rfloor} (-1)^k \binom{\beta}{k} f(t - kh).
\label{eq:gl}
\end{equation}
Similarly, Fig.~\ref{fig:gl_effect} illustrates that the Grünwald-Letnikov derivative is more sensitive to slow-drift anomalies that are characteristic of stealthy cyber-attacks and data manipulation.

The combined feature vector:
\begin{equation}
\mathbf{F}(t) = \left[{}^{C}D_{t}^{\alpha} V, {}^{C}D_{t}^{\alpha} P, {}^{C}D_{t}^{\alpha} Q, {}^{GL}D_{t}^{\beta} V, {}^{GL}D_{t}^{\beta} P, {}^{GL}D_{t}^{\beta} Q\right]^T.
\label{eq:feature_vector}
\end{equation}

Parameters: $\alpha = 0.7$, $\beta = 0.3$, $L = 400$ samples. The impact of these hyper-parameters on validation accuracy is further analyzed in Fig.~\ref{fig:hyperparameter}.

\subsection{Hierarchical Diagnostic Architecture}
\begin{figure}[!t]
\centering
\includegraphics[width=0.45\linewidth]{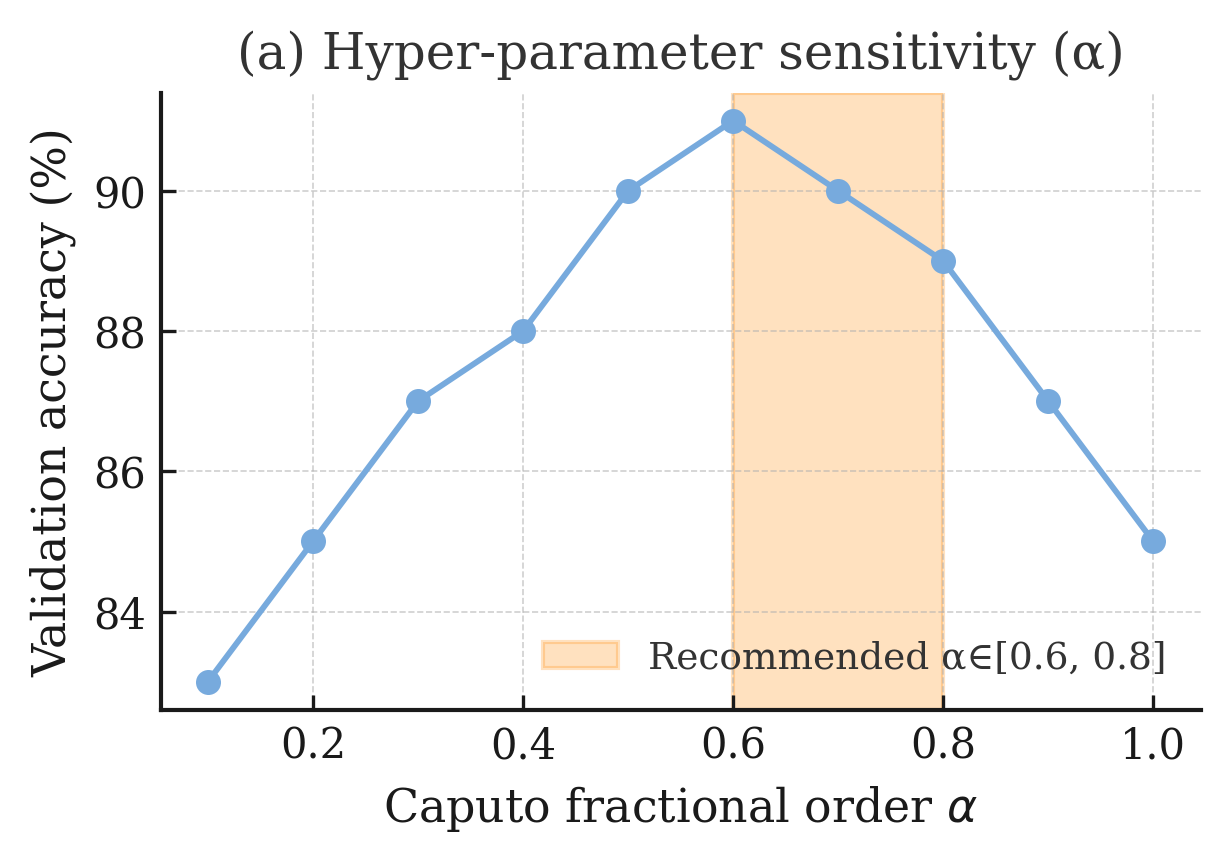}
\hfill
\includegraphics[width=0.45\linewidth]{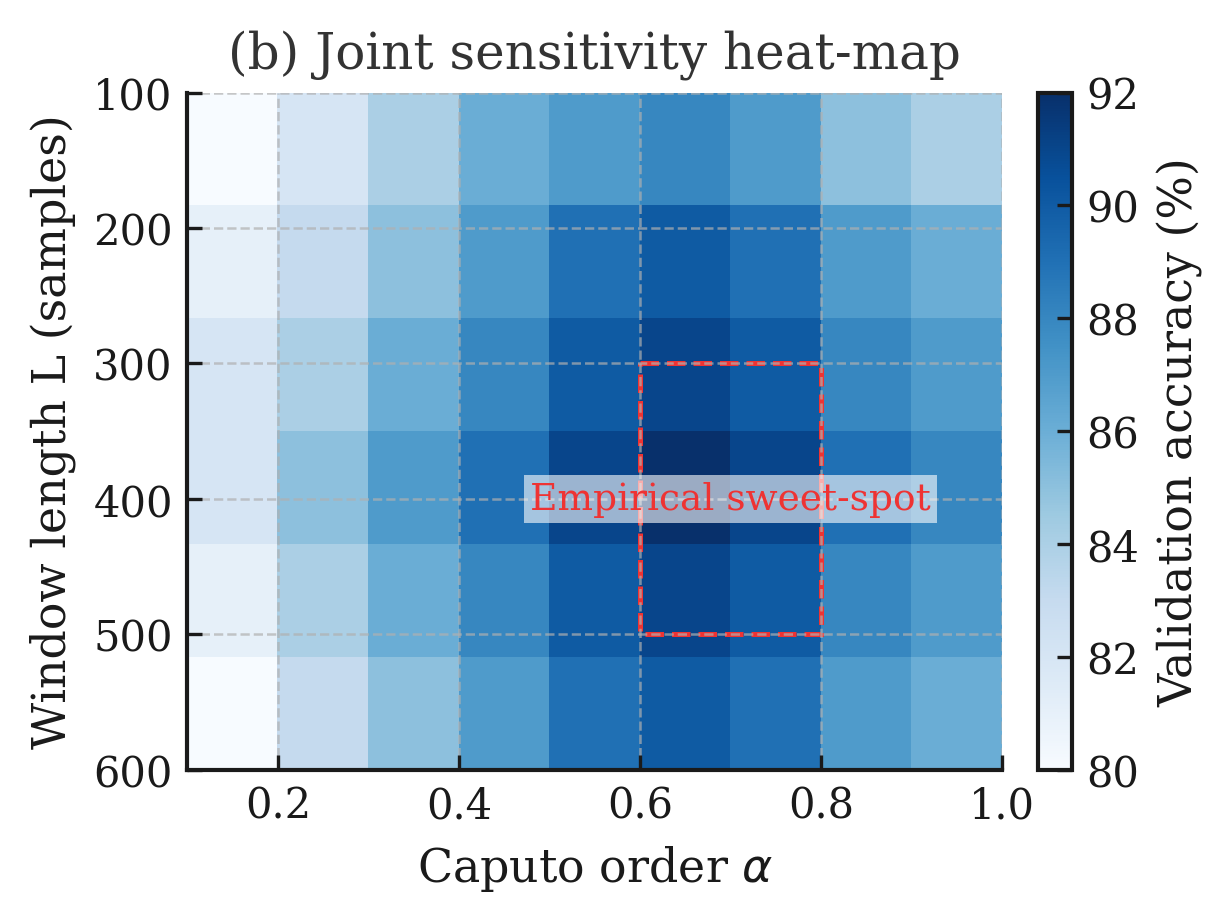}
\caption{Hyper-parameter study of FO-MADS. (a) Validation accuracy versus Caputo fractional order $\alpha$ ($\beta = 0.3$, $L = 400$). (b) Joint sensitivity heat-map of $\alpha$ and window length $L$; darker color denotes higher accuracy. The red rectangle marks the empirical sweet-spot}
\label{fig:hyperparameter}
\end{figure}

Two-stage classification addresses class imbalance: 1) Stage 1: 5-class inverter localization (normal + 4 inverters) and 2) Stage 2: 6-class switch isolation (activated only if a fault is detected). This hierarchical strategy significantly improves switch-level accuracy compared to flat classifiers. The sensitivity of the classifier to the Caputo order $\alpha$ and window length $L$ is summarized in Fig.~\ref{fig:hyperparameter}, which guides the empirical selection of working points.

\subsection{Robustness Training}
Progressive Memory-Replay Adversarial Training (PMR-AT) employs a curriculum of progressively stronger attacks, as sketched in Fig.~\ref{fig:pmr_at}: PGD-based adversarial example generation, Online Hard Example Mining (OHEM), historical attack replay, and progressive attack escalation. The adaptive attack-aware loss weighting schedule used to balance clean and adversarial samples during ablations is shown in Fig.~\ref{fig:adaptive_weight}.

Training curriculum: normal $\rightarrow$ bias $\rightarrow$ noise $\rightarrow$ replacement $\rightarrow$ replay.

\section{Progressive Memory-Replay Adversarial Training}
\begin{figure}[!t]
\centering
\includegraphics[width=\linewidth]{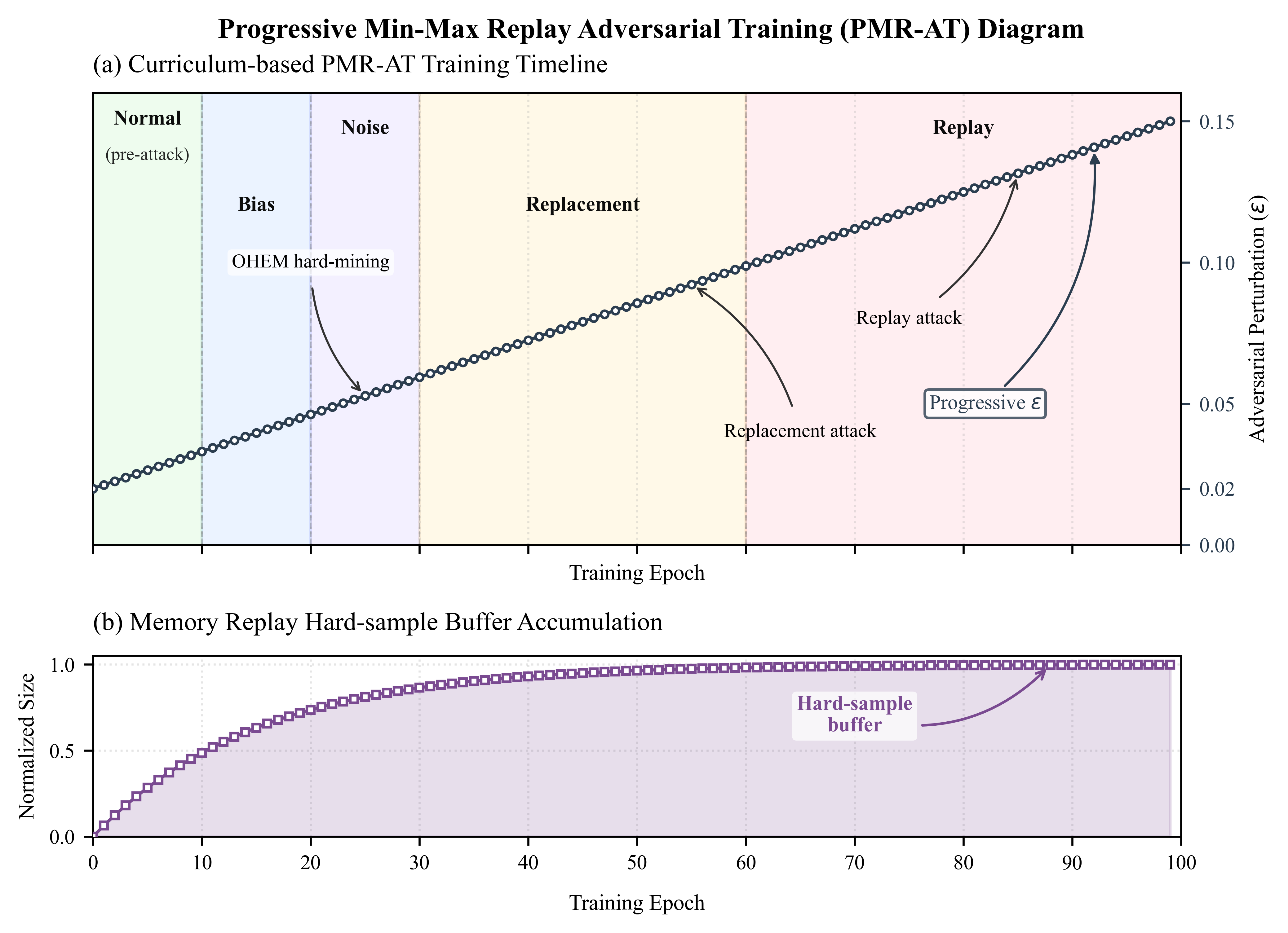}
\caption{Illustration of the PMR-AT training procedure showing progressive attack escalation and memory replay mechanism}
\label{fig:pmr_at}
\end{figure}

\begin{figure}[!t]
\centering
\includegraphics[width=\linewidth]{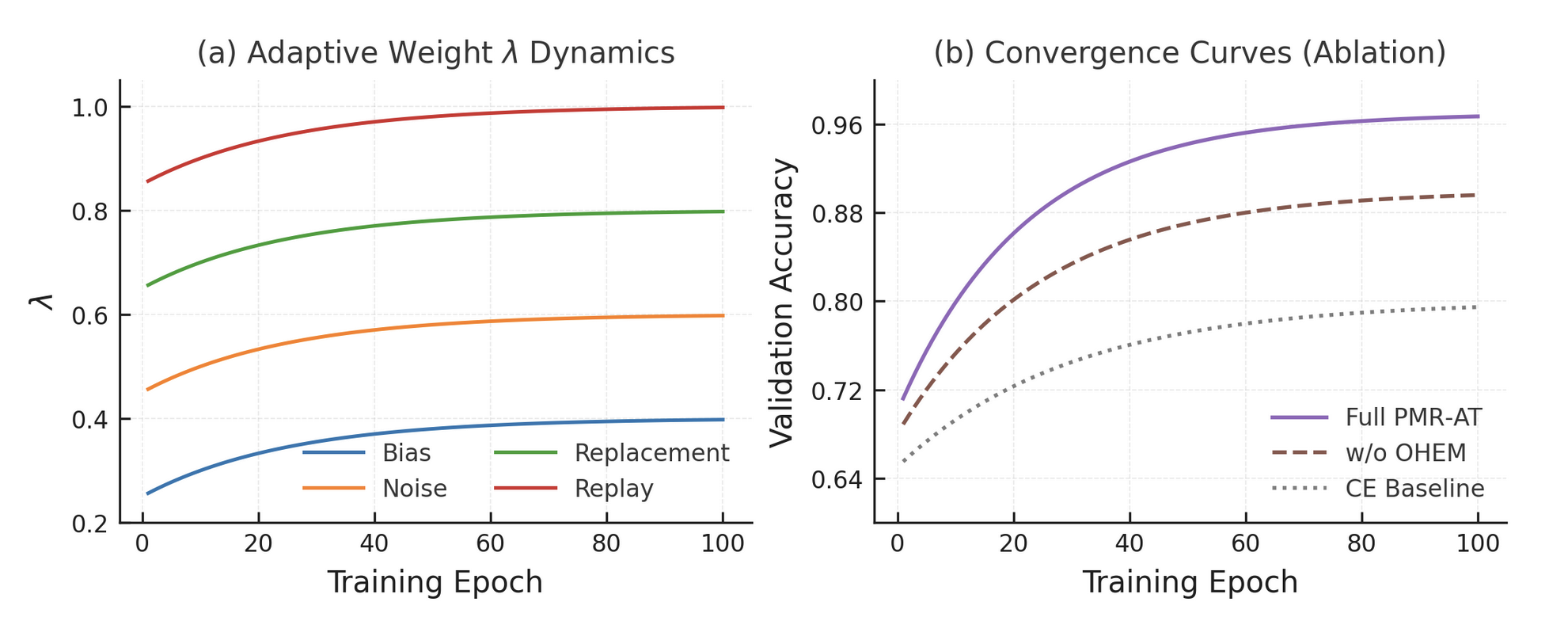}
\caption{Adaptive weight dynamics across components and convergence curves from ablation experiments}
\label{fig:adaptive_weight}
\end{figure}

PMR-AT enhances robustness through 1) multi-stage attack escalation, 2) PGD attacks with $\epsilon=0.1$, 3) historical attack replay, and 4) difficulty progression. The overall pipeline of PMR-AT, including the memory replay buffer and the curriculum of attack stages, is illustrated in Fig.~\ref{fig:pmr_at}, while Fig.~\ref{fig:adaptive_weight} depicts the evolution of the attack-aware weight used in the loss function.

Attack-aware loss function:
\begin{equation}
L_{\text{total}} = L_{\text{CE}}(y, \hat{y}) + \lambda \cdot \mathbb{E}_{(x_h,y_h)\sim \mathcal{D}_{\text{hard}}} [L_{\text{CE}}(y_h, f(x_h))],
\label{eq:loss}
\end{equation}
where OHEM selects top-20\% hard samples, with adaptive weight $\lambda$ scaling with attack difficulty. The weight $\lambda$ is defined as $\lambda = 0.5 \cdot d$, where $d \in [0,1]$ represents the normalized attack difficulty (bias: 0.2, noise: 0.4, replacement: 0.7, replay: 1.0).

\section{Simulation and Experimental Results}
\subsection{Experimental Setup}
\begin{figure}[!t]
\centering
\includegraphics[width=\linewidth]{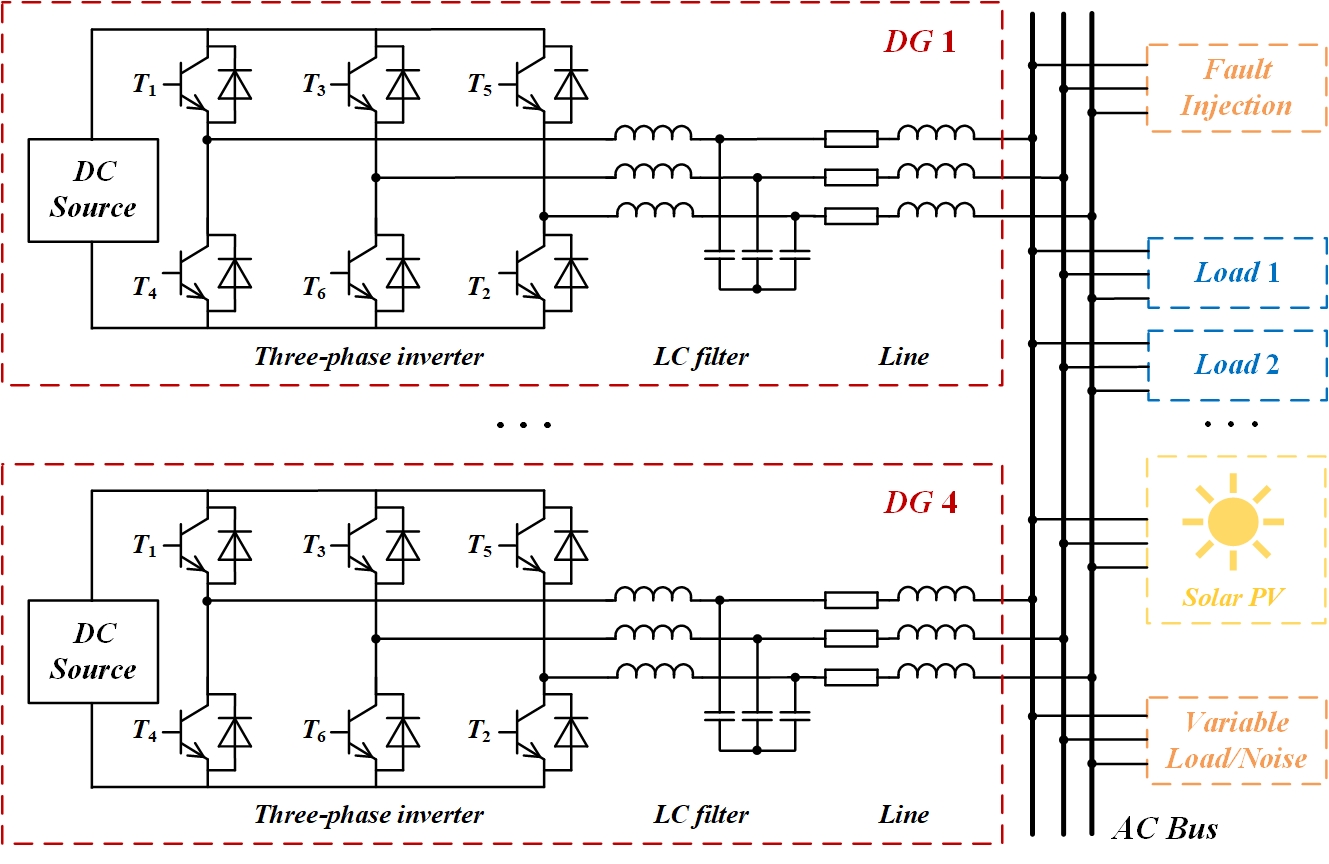}
\caption{Simulink-based four-inverter IBR-dominated microgrid testbed for experimental validation}
\label{fig:testbed}
\end{figure}

A Simulink-based four-inverter testbed generated 5,600 VPQ samples (50 Hz, 2 kHz sampling) with: 24 fault classes (single IGBT open-circuit), 4 attack types (bias, noise, replacement, replay), and an 80/20 train-test split. Bias attacks inject a 10\% DC offset in VPQ, while Gaussian noise attacks use $\sigma = 5\%$ of the nominal magnitude; replacement attacks randomly substitute 20\% of samples with stale measurements. The detailed configuration of the IBR-dominated microgrid and the single VPQ measurement point used for diagnosis are shown in Fig.~\ref{fig:testbed}.

\subsection{Performance Evaluation}
\begin{figure}[!t]
\centering
\includegraphics[width=\linewidth]{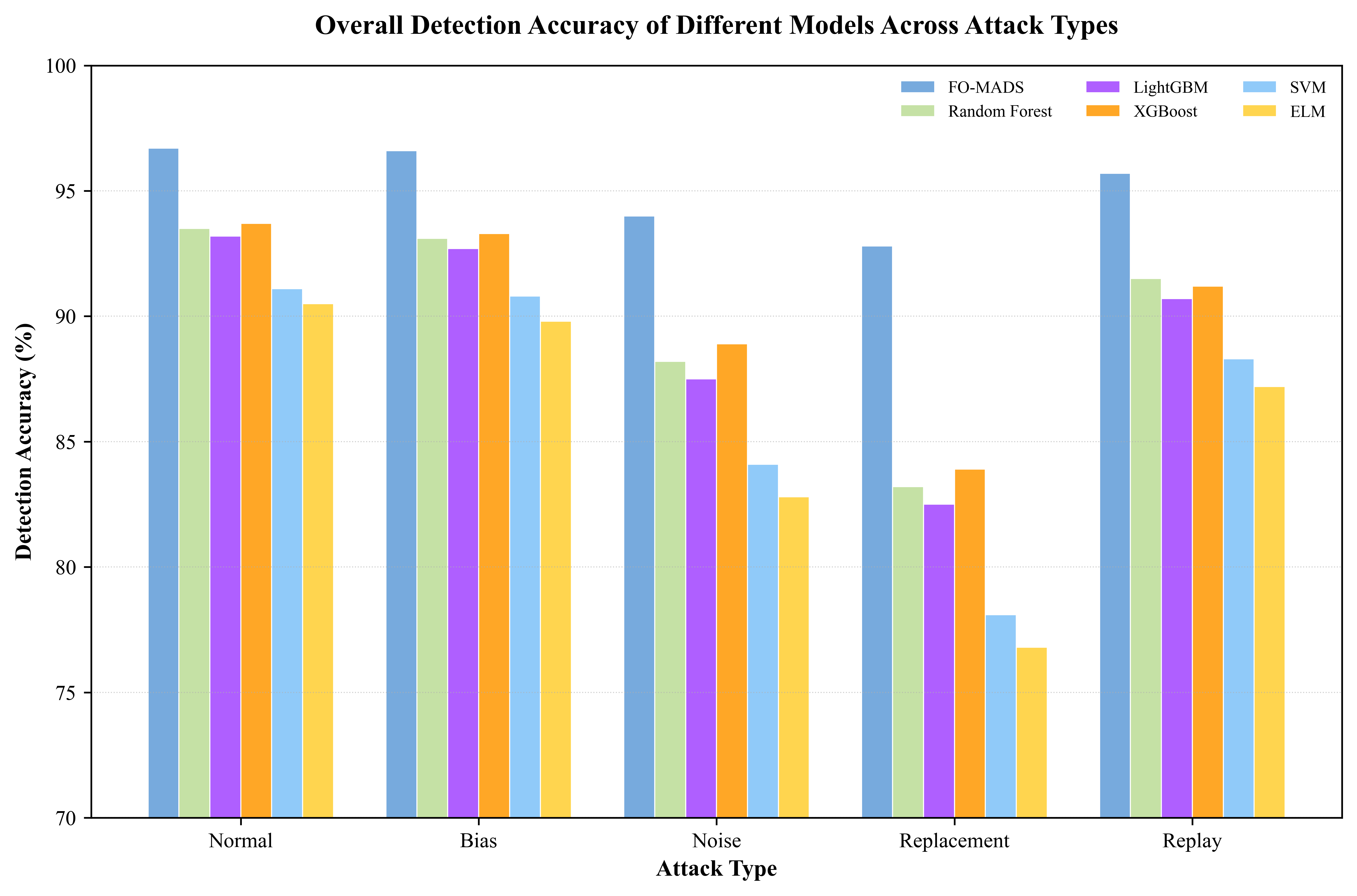}
\caption{Overall detection accuracy of different models across attack types}
\label{fig:model_comparison}
\end{figure}

\begin{figure}[!t]
\centering
\includegraphics[width=\linewidth]{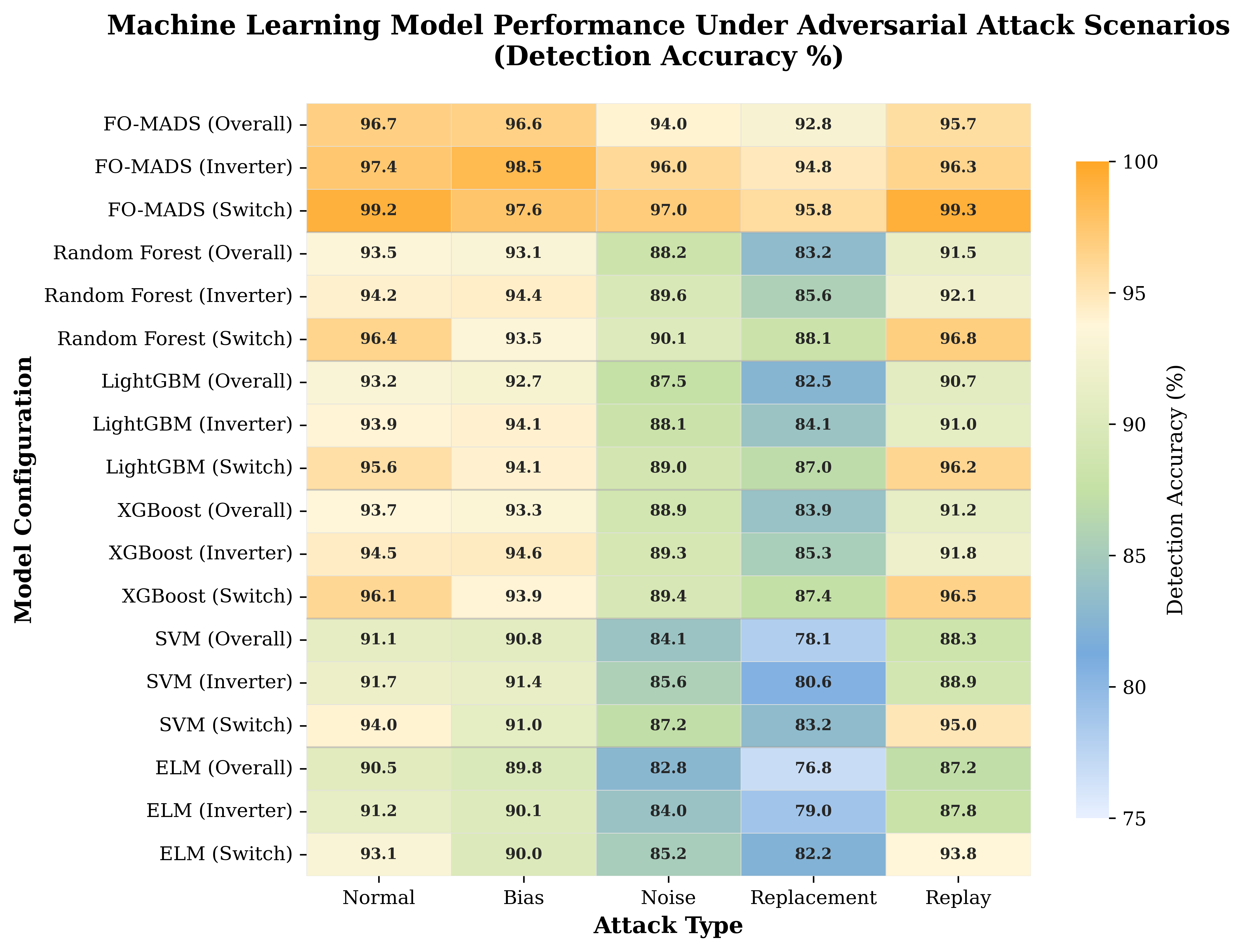}
\caption{Machine learning model performance under adversarial attack scenarios}
\label{fig:performance_heatmap}
\end{figure}

FO-MADS achieves high accuracy across all scenarios, as summarized in Table~\ref{table:overall_accuracy}. A visual comparison of FO-MADS with several baseline models under different attack types is provided in Fig.~\ref{fig:model_comparison}, while Fig.~\ref{fig:performance_heatmap} presents a heat-map view of model performance across attack scenarios.

\begin{table*}[!t]
\caption{OVERALL CLASSIFICATION ACCURACY (\%)}
\label{table:overall_accuracy}
\centering
\begin{tabular}{|l|c|c|c|c|c|c|}
\hline
\textbf{Method} & \textbf{Normal} & \textbf{Bias} & \textbf{Noise} & \textbf{Replacement} & \textbf{Replay} & \textbf{Remarks} \\ \hline
FO-MADS (Proposed) & 96.7 & 96.6 & 94.0 & 92.8 & 95.7 & Full implementation with all components \\ \hline
FO-MADS w/o OHEM & 95.1 & 94.3 & 90.2 & 87.5 & 92.0 & Without Online Hard Example Mining \\ \hline
FO-MADS w/o Frac. Feat. & 93.8 & 92.6 & 88.1 & 85.3 & 90.4 & Without fractional-order features \\ \hline
XGBoost & 93.7 & 93.3 & 88.9 & 83.9 & 91.2 & Baseline model 1 \\ \hline
Random Forest & 93.5 & 93.1 & 88.2 & 83.2 & 91.5 & Baseline model 2 \\ \hline
CNN & 92.0 & 90.5 & 85.8 & 80.1 & 88.9 & Baseline model 3 \\ \hline
\end{tabular}
\begin{center}
\footnotesize
\end{center}
\end{table*}

Hierarchical breakdown is reported in Tables~\ref{table:inverter_accuracy} and \ref{table:switch_accuracy}. Specifically, inverter localization remains above 94.8\% under all attack types, as shown in Table~\ref{table:inverter_accuracy}, while switch isolation maintains above 95.8\% accuracy even in adversarial cases, as summarized in Table~\ref{table:switch_accuracy}.

\begin{table}[!t]
\caption{INVERTER-LEVEL ACCURACY (\%)}
\label{table:inverter_accuracy}
\centering
\begin{tabular}{|l|c|c|c|c|c|}
\hline
\textbf{Attack Type} & \textbf{Normal} & \textbf{Bias} & \textbf{Noise} & \textbf{Replacement} & \textbf{Replay} \\ \hline
FO-MADS & 97.4 & 98.5 & 96.0 & 94.8 & 96.3 \\ \hline
\end{tabular}
\end{table}

\begin{table}[!t]
\caption{SWITCH-LEVEL ACCURACY (\%)}
\label{table:switch_accuracy}
\centering
\begin{tabular}{|l|c|c|c|c|c|}
\hline
\textbf{Attack Type} & \textbf{Normal} & \textbf{Bias} & \textbf{Noise} & \textbf{Replacement} & \textbf{Replay} \\ \hline
FO-MADS & 99.2 & 97.6 & 97.0 & 95.8 & 99.3 \\ \hline
\end{tabular}
\end{table}

\begin{figure}[!t]
\centering
\includegraphics[width=\linewidth]{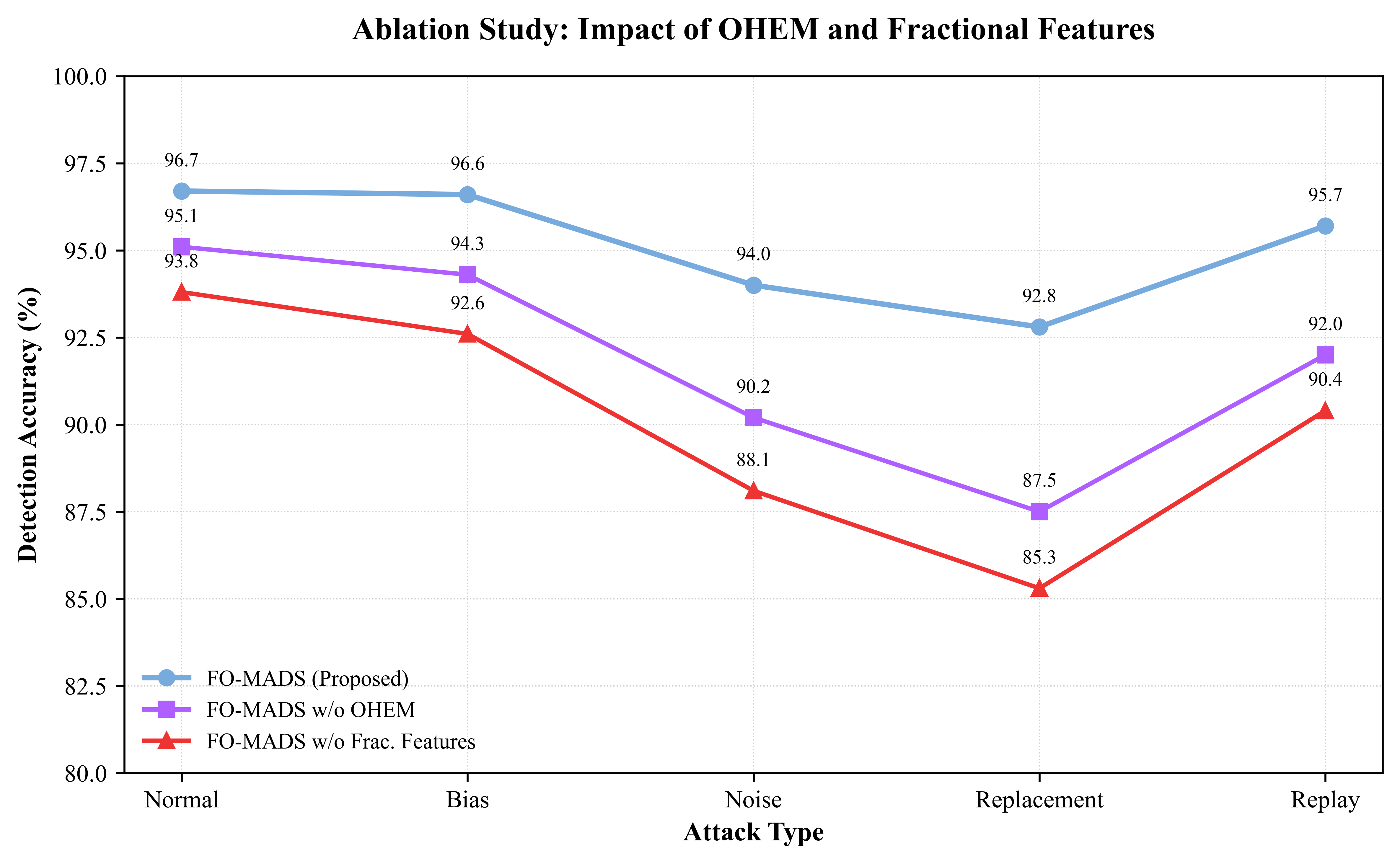}
\caption{Ablation study impact of OHEM and fractional features}
\label{fig:ablation_study}
\end{figure}

Ablation analysis confirms the benefit of the proposed components (see Fig.~\ref{fig:ablation_study}): dual fractional features boost noise and replacement robustness by 5.9\% and 7.5\%, respectively; OHEM further reduces switch-level misclassification under noise; and PMR-AT enhances replay-attack resilience. The corresponding attack-wise model comparison and scenario-wise performance, including additional baselines (LightGBM, SVM, and ELM) beyond those summarized in Table~\ref{table:overall_accuracy}, are depicted in Fig.~\ref{fig:model_comparison} and Fig.~\ref{fig:performance_heatmap}.

\section{Conclusions}
This study introduced FO-MADS, a cost-effective framework achieving cyber-physical resilience for inverter-based resource-dominated microgrids using only a single VPQ sensor. By exploiting Caputo and Grünwald-Letnikov derivatives, FO-MADS constructs a dual fractional-order feature library that magnifies both high-frequency perturbations and slow-drift anomalies. A two-stage hierarchical classifier localizes the faulty inverter and isolates the defective IGBT switch, while PMR-AT systematically hardens the model against cyber-attacks. Extensive simulations yielded 96.7\% accuracy under attack-free operation and above 92.8\% across all attack scenarios.

The main contributions include: 1) single-sensor diagnosis eliminating multi-point instrumentation; 2) dual-definition feature engineering using complementary fractional operators; 3) hierarchical localization significantly improving switch-level accuracy; and 4) cyber-resilient fault diagnosis with PMR-AT that significantly improves adversarial accuracy across diverse attack types.

Future work will focus on hardware-in-the-loop trials, embedded implementations, extended fractional-order features, and distributed privacy-preserving variants, advancing FO-MADS toward fully deployable real-time solutions for resilient power distribution networks.

\end{document}